\documentclass[twocolumn]{article}
\usepackage{amsmath}
\usepackage{amssymb}
\usepackage{graphicx}
\usepackage{dcolumn}
\usepackage{bm}
\usepackage{soul}
\usepackage{comment}
\usepackage{epstopdf}
\usepackage{float}
\usepackage{color}
\usepackage{soul}
\usepackage{changes}

\newcommand{\Eqref}[1]
{equation (\ref{#1})}

\newcommand{\Figref}[1]
{Fig. \ref{#1}}

\newcommand\Name[1]{#1}
\newcommand\name[1]{#1}
\newcommand\REVIEW[1]{#1}
\newcommand\Book[1]{#1}

\title{\LARGE
Extruding the vortex lattice: two reacting populations of dislocations}

\author{J.S. Watkins \and N.K. Wilkin}
\date{%
	 School of Physics and Astronomy, University of Birmingham, Birmingham, B15 2TT, UK\\[2ex]%
	\today
}

\begin{document}

\maketitle
\section*{Abstract}
{\bf A controllable soft solid is realised in vortex matter\cite{eskildsen2011,guillamon2014,lukyanchuk2015} in a type II superconductor. The two-dimensional unit cell area can be varied\cite{fasano2008} by a factor of $10^4$ in the solid phase, without a change of crystal symmetry offering easy exploration of extreme regimes compared to  ordinary materials. The capacity to confine two-dimensional vortex matter to mesoscopic regions\cite{kes2004,lukyanchuk2015} provides an arena for the largely unexplored metallurgy of plastic deformation {\em at large density gradients}. Our simulations reveal a novel plastic flow mechanism in this driven non-equilibrium system, utilising {\em two} distinct, but strongly interacting, populations of dislocations. One population facilitates the relaxation of density; a second aids the relaxation of shear stresses concentrated at the boundaries. The disparity of the bulk and shear moduli in vortex matter ensures the dislocation motion follows the overall continuum flow reflecting density variation.}

Soft matter forms a versatile laboratory to study plastic deformation, including: the observation of dislocation nucleation\cite{spaepen1}, motion \cite{spaepen2, hirth2009,kukubo2002,kes2004,miguel2011,ray2014}, reactions\cite{Irvine} and role in grain boundary processes \cite{menezes}. Soft {\em vortex} matter has the specific advantage that the density of vortices can be changed easily by altering the magnetic field applied, and a density {\em gradient} is created by applying a field gradient \cite{ray,irvine2}. The regime of large density gradients has been extensively studied in colloidal systems \cite{koppl2006, peeters2009}. Here, the regime of large density gradients in vortex matter is naturally studied by extrusion along a channel between reservoirs of different densities. The resulting time dependent non-equilibrium state is the subject of this article.

\begin{figure*}
\includegraphics[width=\textwidth]{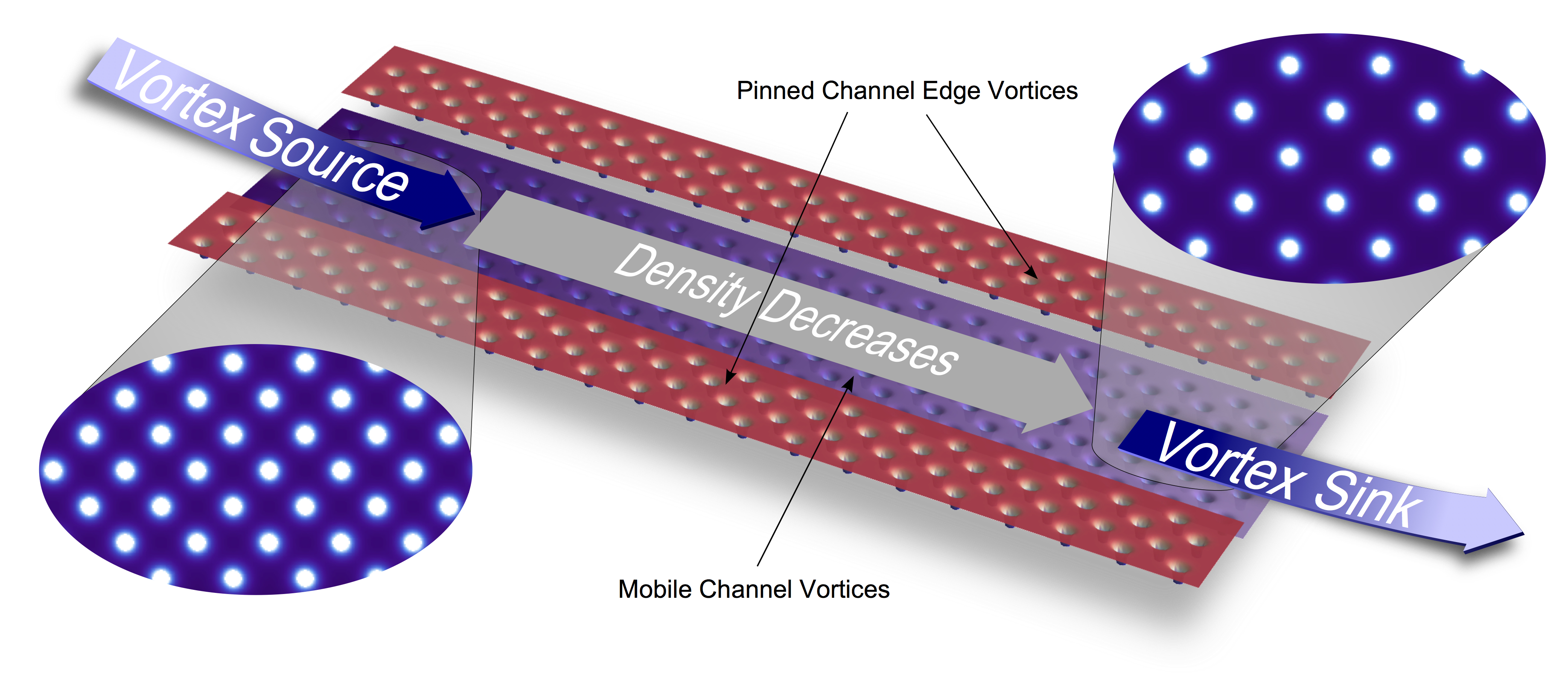}
\caption{\label{fig:fig_1}{\bf The model of a thin channel superconductor with an imposed magnetic field gradient} $B(x) \hat{\bf z}$, where $z$ is normal to the plane. Mobile vortices from a high-density source move along the channel under the action of a vortex density gradient. The channel edges are defined by pinned vortices.}\end{figure*}

The channel is formed by a clean (unpinned) region of width $w$ between walls provided by two pinned regions of vortex lattice. Altering the external magnetic field alters the density of vortices within the channel, while the pinned regions are unaltered for moderate changes of field.  Except when we explicitly compare with the liquid phase, our simulations are at a sufficiently low temperature that--for our finite sample--there are no thermally excited Halperin-Nelson-Young dislocations.

That vortex dynamics is collective in such a channel was demonstrated\cite{kes2004} by the application of the electrical current to a small region of the channel which generated motion of vortices up to 5$\mu{\rm{m}}=30w$ away. This implies a value of 5$\mu$m for the Larkin-Ovchinnikov length\cite{larkin1979}, over which the vortex lattice is not pinned. Motivated by these results, we will consider the clean limit for the channel in this article, with an ordered pinned lattice defining the channel edges.

To investigate flow (both in solid and liquid phases) at controllable density gradients, our simulations add a reservoir with a chosen vortex density to each end of the channel (\Figref{fig:fig_1}). Experimentally, the reservoirs could be fed via vortex pumps\cite{cole2006}; in the simulation vortices are added or removed sufficiently remotely from the channel exit and entrance so as not to affect the flow. This is achieved by calculating the field in each reservoir and adding or removing vortices at the lateral edges of each reservoir in order to maintain the required field. Using this method gives more control over the field gradient than a periodic boundary condition on the flow.

The geometry of the channel is shown in \Figref{fig:fig_1}, where $B_{\rm L}$ and $B_{\rm R}$ are the fields in the left and right reservoirs, with $B_{\rm L}>B_{\rm R}$, favouring vortex motion from left to right in the channel.  We work in the regime where the average density in the channel is comparable to the pinned lattice, so experimental changes of field would be small.  We examine a ``wide" channel of width, $w\sim10 a_0$, where $a_0$ is the lattice parameter of the pinned lattice, which is our unit of length (and the associated unit of field, $B_0$).  So, although the channel lattice is only slightly mismatched with the pinned lattice, the cumulative effect {\em across} the width of the channel can be several lattice parameters. The ``wide" channel will allow a continuum description.

\begin{figure}
\includegraphics[width=\columnwidth]{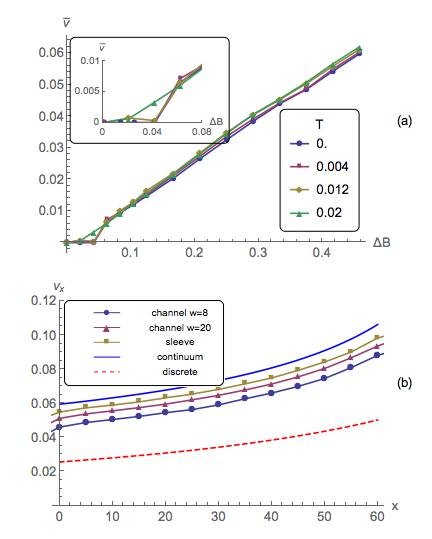}
\caption{\label{fig:fig_2} {\bf Overall flow, yield and acceleration along channel.} {\bf a} shows the variation of the vortices' average velocity, $\overline v(\Delta B)$,  with field difference. There is a critical field difference for the solid to yield at sufficiently low temperatures, which disappears above the melting temperature (similar to the velocity/Lorentz relation from the Leiden group\cite{kukubo2002,kes2004}). {\bf b} shows the velocity profile, $v_x(x)$, in the channel and sleeve (cylindrical) geometries, with $\Delta B=0.46$. To provide bounds on the velocity profiles, we also show continuum calculations for $v(x)$ and a cutoff nearest-neighbour only discrete lattice sum. As the channel width grows $v(x)$ approaches the cylindrical result (which is closer to the continuum model), showing the diminishing effect of edge shear.
}
\end{figure}

\Figref{fig:fig_2}a shows the yield stress for plastic flow at $B_{\rm L}-B_{\rm R}=\Delta{B}_{\rm y}=0.08$ for $T=0$.  Above the yield stress ${\overline v}\propto{(\Delta{B})}$, i.e. linear to a good approximation. In the liquid phase, for $T>T_{\rm m}=0.014$, linearity is present for all $\Delta{B}$. That $\Delta{B}_{\rm y}$ and $T_{\rm m}$ are numerically small reflects the disparity of bulk and shear moduli in the vortex lattice.

\begin{figure}
\includegraphics[width=\columnwidth]{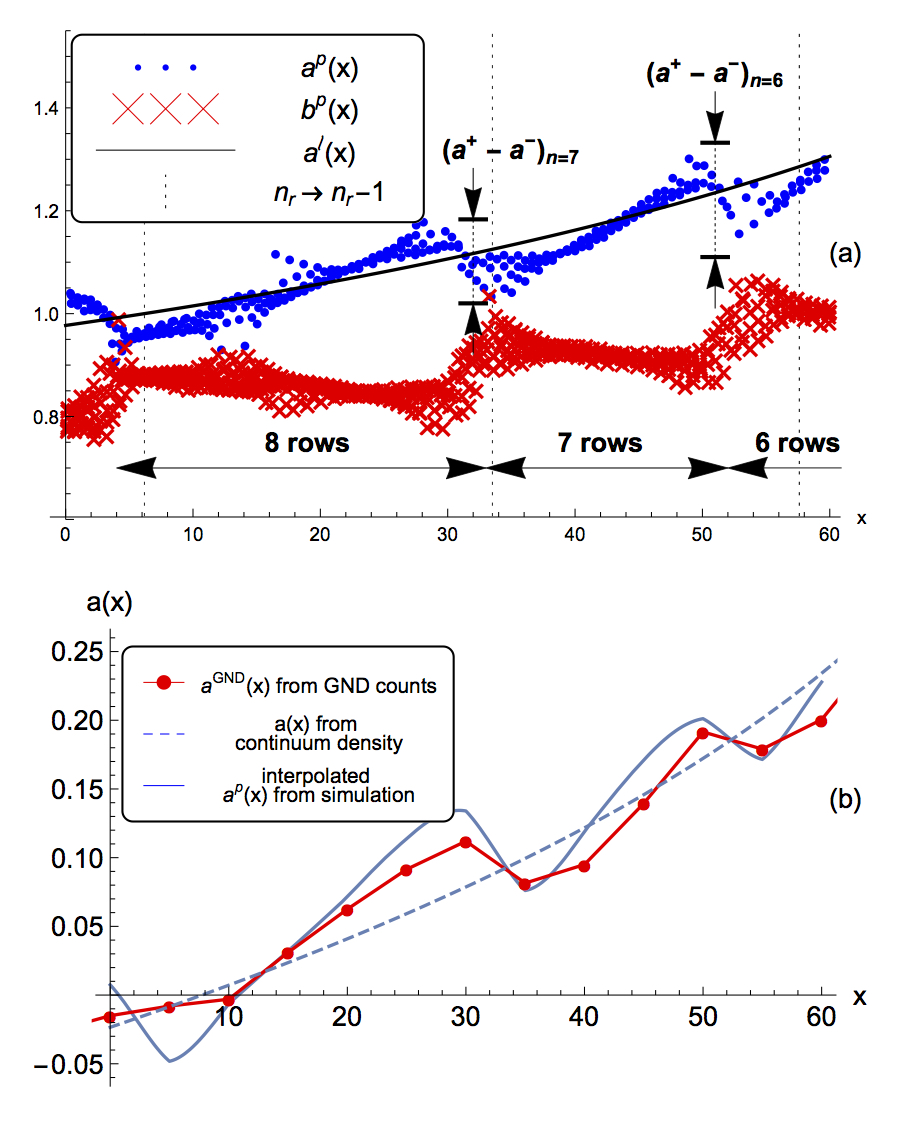}
\caption{\label{fig:fig_3} {\bf The discontinuous evolution of the lattice along the channel.} {\bf a} shows the variation in the vortex spacing (found using Delaunay triangulation) parallel to the channel boundary, $a(x)$, with vertical arrows indicating jumps mentioned in the text. The component of the separation {\em perpendicular} to the channel edges, $b(x)$, is also plotted.
The system contains three zones of $n_r=8$, 7 and 6 rows of vortices.  {\bf b} shows the density of GNDs. The solid line is calculated using an interpolated $a^{\rm p}(x)$ from the simulation and \Eqref{eq:rhog}. The dashed line is a continuum prediction. The red line is from the simulations.}
\end{figure}

A reference for density changes along the channel is provided by the local vortex spacing in the {\em liquid} phase, $a^\ell(x)$, which is smooth:
$$a^\ell(x)\simeq\sqrt{\frac{2}{\sqrt{3}}\dfrac{\Phi_0}{(B_{\rm R}-B_{\rm L})(x/L)+B_{\rm L}}},$$
where $L$ is the channel length and $\Phi_0$ is the flux quantum. If the ``solid", plastic, phase were glassy or hexatic, the density might vary continuously as well. However, as can be seen from \Figref{fig:fig_3}a, this is not true. While the inter-vortex spacing {\em parallel} to the channel, $a^{\rm p}(x)$, tracks the liquid variation, $a^\ell(x)$, the {\em perpendicular component} of the spacing, $b^{\rm p}(x)$ ($b=\sqrt{3}/2a$ for an equilateral triangular lattice), is step-like along the channel.

\begin{figure*}
\includegraphics[width=\textwidth]{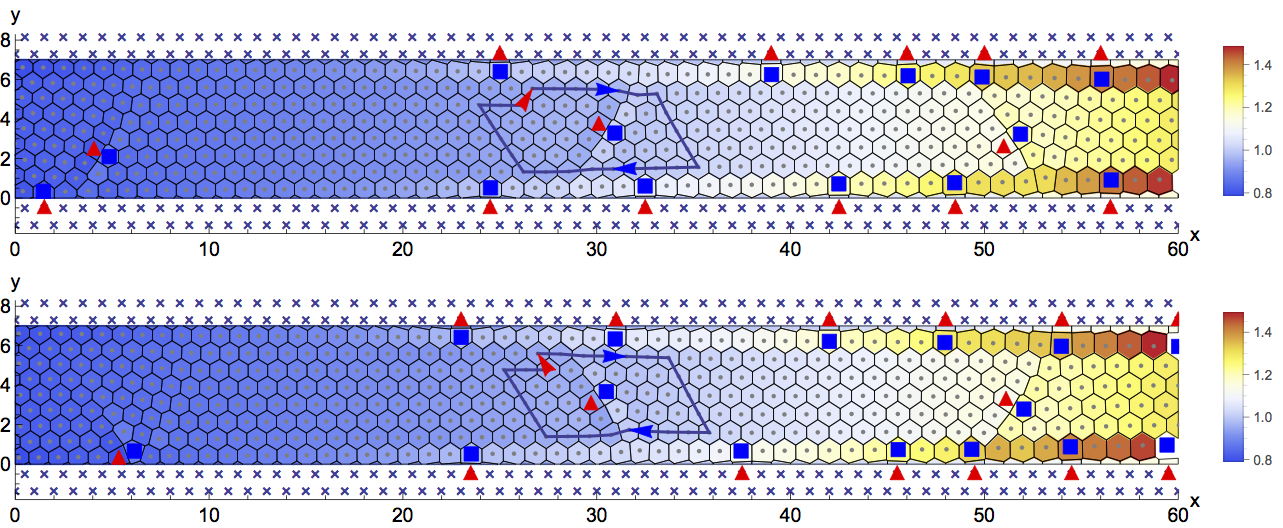}
\caption{\label{fig:fig_4} {\bf The double dislocation network in the channel.}  A snapshot of the vortex positions in a channel, of width $w=8b_0$, is shown. There are three ``bulk" edge dislocations. The Burgers circuit construction\cite{hull2001} for the second dislocation is indicated. Pinned vortices in the channel edges are marked with ($\times$). $\triangle$ and $\Box$ mark vortices with $5$ and $7$ neighbours respectively. All other vortices have 6 neighbours.  (Top) An bulk dislocation approaching the lower channel edge. \, (Bottom) The same dislocation after the interaction with a GND with ${\bf b}=-\hat{\bf y}$.}
\end{figure*}

The interpretation, confirmed by examination of \Figref{fig:fig_4}, is that the vortex matter is mostly crystalline with the inter-row spacing commensurable with the channel width. The commensurability dictates discrete changes along the channel, where rows disappear, associated with an edge dislocation in the ``bulk" of the channel. Because vortex matter has no cohesive energy, the inter-row separation expands (and the unit cell expands) as $x$ passes an edge dislocation, the lattice filling the channel laterally with fewer rows. The required number of bulk dislocations is increased by increasing the magnetic field gradient or the width of the channel (which requires more rows to be removed for a given density change). Our simulations demonstrate this for density gradients necessitating up to 4 edge dislocations, with channel widths of up to $30b_0$.

The unit cell changes shape from a compressed isosceles triangle to an equilateral triangle upon passing an edge dislocation. I.e. the transition from $n+1$ rows to $n$ rows occurs when $b=(\sqrt{3}/2)a^{\rm p}(x)=w/n$. To avoid gross mechanical disequilibrium, we expect the unit cell {\em area} to be continuous as a function of $x$. Equating the unit cell sizes in the sections with different rows at the boundary implies a discontinuity in $a^{\rm p}(x)$,  $a^{+}-a^{-}=(2/\sqrt{3})(w/n^2)$, where $a^+$ is the lattice parameter on the side with $n+1$ rows and $a^-$ that with $n$ rows. This difference is indicated in \Figref{fig:fig_3}a, agreeing with the simulations.

The ``geometrically necessary strain" caused by the lattice parameters of the pinned region and the channel lattice becoming increasingly mismatched as $x$ increases is concentrated in ``misfit", or geometrically necessary dislocations (GNDs) at the interface \Figref{fig:fig_4}).  The ``charge" density of GNDs, $\rho_{\rm g}$, reflecting the lack of registry due to the variation in $a^{\rm p}(x)$, is:
\begin{equation}
\rho_{\rm g}=\frac{1}{a_0}\left(
1-\dfrac{a_0}{a^{\rm p}(x)}\right)\label{eq:rhog}
\end{equation}

\Figref{fig:fig_3}b shows the agreement between this expression and the density of GNDs found in the simulation.

The {\em dynamic} behaviour (see Supplementary Video S1) of the plastic flow reflects the {\em interacting} populations of GND and ``bulk" dislocations.  The GNDs glide parallel to the channel edges, lubricating the vortex lattice motion along the channel. The bulk dislocations glide on symmetry-related glide-planes across the channel.  The video {\em appears} to show that bulk dislocations are reflected at the channel edge onto to the other glide-plane not parallel to the channel edge, and repeat this zig-zagging motion between the channel edges. We have followed this periodic motion for more than 100 periods.

However, it {\em{cannot}} be a reflection, as the {\em{conserved}}\cite{hull2001} Burgers vector changes when gliding on different planes. The resolution is that a ``reaction" occurs, visible in Supplementary Video S1: a bulk dislocation upon reaching a channel edge combines with a GND producing an bulk dislocation on the third glide-plane (i.e. the three possible Burgers vectors add to zero).

The steady state of plastic flow is constituted by the regions of constant row number, delimited, in the laboratory-frame, by the average $x$-coordinates of the zig-zagging bulk dislocations. The gliding GND dislocations ensure this.

Building a global picture of the flow down the channel from these local descriptions of dislocation motion is aided by \Figref{fig:fig_2}a. Note the near identity of flow rates in liquid and plastic phases--despite the considerable difference in structure. The underlying cause is that vortex matter is soft but incompressible\cite{Brandt1995}: the ratio of the bulk, $\kappa$, to shear, $\mu$, moduli is $\kappa/\mu=16\pi(\lambda/\xi)^2\gg{1}$, for strongly type II superconductor, where $\xi$ is the coherence length and $\lambda$ is the penetration depth of the superconductor. Thus the macroscopic flow rate, reflecting density gradients, is insensitive to crystalline order and the steady-state profile for $v(x)$ and $\rho(x)$ along the channel may be derived using the continuity equation for the vortices and the force equation on each vortex (see Supplementary Material),
$$v(x)=-\frac{\Phi_0^2}{\eta\pi\mu_0}\thinspace\frac{{\rm d}\rho}{{\rm d}x}$$
They yield:
$$\rho (x)=\rho(0)\sqrt{1-\frac{x}{L_0}};\quad{v(x)}=\frac{Q}{\rho(x)},$$
where $x =0$ has been chosen to be the start of the channel, $Q=\rho(x)v(x)$ is conserved in steady state and $L_0=\Phi_0^2\rho(0)^2/(\pi\eta{Q}\mu_0)\gg{L}$ in our simulations (i.e. the number of rows does not drop to zero).  The resulting velocity field is shown in \Figref{fig:fig_2}b.

The microscopic dislocation motion is slaved to this density--gradient dominated continuum description (i.e. determined kinematically) as the Peierls-Nabarro stress for glide is determined\cite{hull2001} by the small shear modulus. The GNDs ensure the average motion of the channel lattice occurs with the velocity $v(x)$: each GND translates the lattice by $a_0$ as it passes, so their velocity, $v_{\rm g}(x)=v(x)/(a_0\rho_{\rm  g}(x))$.

\begin{figure}
\includegraphics[width=\columnwidth]{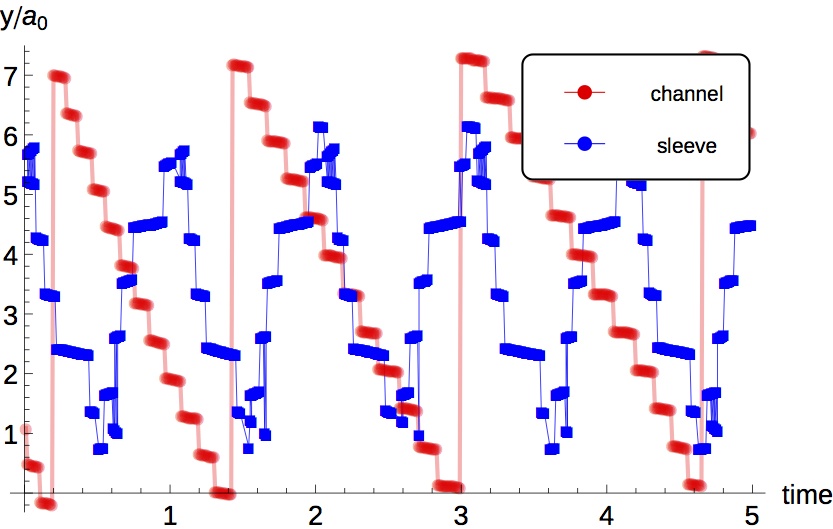}
\caption{\label{fig:fig_5} {\bf Motion of the ``bulk'' dislocations, kinematic and dynamic.} Dislocation paths are shown for both the channel and the sleeve, in a reference frame moving with the channel/sleeve lattice. In the case of the channel, non-kinematic influences are implied an additional velocity modulation: moving faster as they leave a channel edge and slowing are they approach an edge. This is due to image forces being repulsive due to the rigid pinned lattice, although this is partly cancelled by the lubrication of the GND's allowing slip along the surface\cite{Head1953}. There is no correlation between the different zig-zagging dislocations--presumably because their velocities are different (as $v(x)$ varies) and interactions are suppressed by exponential screening due to the image arrays.
}
\end{figure}

The zig-zagging dislocations ensure that the density profile is stationary in the laboratory frame. They move backwards, see \Figref{fig:fig_5}, at an average velocity $v_{\rm zig}=-2v(x)$, where the factor of two comes from the angle of the glide plane.  Channel-edge friction can be removed by considering a ``sleeve", with a periodic boundary condition in the $y$-direction. On the sleeve there are still preferred row separations due to commensurability with the circumference of the cylinder. \Figref{fig:fig_2}b shows indeed that the sleeve-system is closer than the channel to the continuum model. This is then reminiscent to the description\cite{nelson2012,amir2014} of bacterial cell wall growth and provides a physical mechanism for the observations in colloidal dynamics as seen in Deutschl\"{a}nder \emph{et al.}\cite{deutschlaender2013}.

In summary, the first study of plastic deformation under significant density gradients has demonstrated the existence of a new steady-state with a strongly interacting set of dislocations on all of the glide planes of the vortex crystal. Whilst the vortex crystal has no cohesive energy, one would expect similar behaviour for any two-dimensional matter compressed sufficiently from its equilibrium density. The generalization to three dimensions--either for flux lines or particles--is an open question, as is the potential of the latter for high/low compressibility (cf pinned/channel) heterogeneous mixtures in geophysical flows. 

\section*{Acknowlegdements}
It is a pleasure to acknowledge stimulating discussions with Mike Gunn. The authors acknowledge financial support from the EPSRC for student funding. The authors declare no competing financial interests.

\section*{Author Contributions}
J. S. Watkins and N. K. Wilkin wrote the main manuscript text and prepared figures 1-5. All authors reviewed the manuscript.


\section*{Competing financial interests}
The authors declare no competing financial interests.

\section*{Methods}
\textbf{Simulation Techniques.}\\ The motion of the $N$ two dimensional vortices is represented via molecular dynamics, describing the vortices as particles with repulsive interactions and following the diffusion dynamics of Jensen {\em et. al.}\cite{jensen1993,reichhardt2001}:
$$\eta {\bf v}_i = {\bf F}^{\rm vv}_i + {\bf F}^T_i, \label{eq:eqofmotion}$$
${\bf v}_i$ is the velocity of the $i$th vortex with $\eta$ an effective viscosity due to the normal fluid. Temperature is included via ${\bf F}^T_i$, a thermostat \cite{jensen1993,dong1993} with $\langle  F^T_i \rangle=0$ and $\langle F^T_i(t) F^T_j(t')\rangle =2 k_B \eta T \delta_{ij} \delta(t-t')$. Finally, the vortices interact via the standard\cite{besseling2005} repulsive force:
$${\bf f}^{\rm vv} ({\bf r}) = - f_0 K_1 \left( r/\lambda \right) {\hat {\bf r}}; \quad {\bf F}^{\rm vv}_i = \sum_{j=1\ne i}^N {\bf f}^{\rm vv} ({\bf r}_i-{\bf r}_j) \label{eq:force} $$

where $ f_0 = \Phi^2 / 2 \pi \mu_0 \lambda^3 $,  $\lambda$  is the  penetration depth, $\Phi$ is the flux quantum, $K_1$ is a modified Bessel function and ${\bf F}^{\rm vv}_i$ is the total force on the $i$th vortex due to the others. Reasonable values of the pinned vortices lattice parameter are chosen based on experiment\cite{pruymboom1988}. We use the lattice parameter, $a_0 = 100 \textrm{nm}$, of the pinned lattice as the unit of length, fixing the penetration depth $\lambda =1.11 a_0$ followings the values used by Besseling \emph{et  al.} and ensures a separation of bulk and shear moduli; $C_{11}\gg C_{66}$. Magnetic fields are described in units of the pinned vortices field which we take to be $B_0=0.24 \textrm{T}$.  For simulation purposes we use a force cut-off range\cite{reichhardt2001} set at $r_{\rm{cut}}=6 \lambda$. We let $k_B = \eta = f_0 = 1$. (This choice of units gives a fundamental mass of $M = \eta a_0/f_0 = 1$ and time $T = \eta^2 a_0/f_0 = 1$.) The magnetic fields at the ends of the channel are maintained via (large) vortex reservoirs, \Figref{fig:fig_1}, which are held at the required fields by the addition or removal of vortices.

Following the method of Spencer \emph{et al.} \cite{spencer1997}, density of defects and the rotational order parameter were used to determine a melting temperature for the bulk system as $T_{\rm m}=0.014$.

Our simulations are almost all deep in the solid phase, well below $T_{\rm m}$ except when we make comparison with the liquid phase and show the absence of a yield stress for $T>T_{\rm m}$. We confirm the solid nature by structure factor measurements\cite{spencer1997}.  The lattice parameter of the pinned vortices at the channel edge (CE) was $a_0$ corresponding to $B=1$ and row spacing  $b_0= \sqrt{3}/2 $. The majority of runs fixed $B_{\rm L}=1.05$ such that the lattice parameters of the CE and flowing vortices coincide at $x=6.17$ along the channel. The channel length for all simulations was set at 60.

The time step for the simulations was chosen\cite{besseling2005} to be $\delta t=0.01$ to ensure the maximum vortex displacement in one iteration was $\lesssim a_0/50$. Results are taken after at least 100 000 time steps, at which time the vortices had reached a non-equilibrium steady-state.

\noindent
\textbf{Definition of geometrically necessary dislocation density}
Note this {\protect \em continuum} relation
$$\rho_{\rm g}=\frac{1}{a_0}\left(
1-\dfrac{a_0}{a^{\rm p}(x)}\right)\label{rhog}$$
breaks down when $\rho _{\protect \rm g}(x) \to 0$, as, in the context of this article, it happens over a finite region, whereas the continuum description is appropriate on scales, $\lambda $, which are large compared to the (divergent, as $\rho _{\protect \rm g}(x) \to 0$) inter-dislocation separation--which is a contradiction.

\noindent
\textbf{Continuum analysis.} Modelling the system using the continuum approximations allows for the spatial variation of $\rho(x)$ and $v(x)$ to be determined. The starting point is to ignore edge effects, working more than a penetration depth into the channel, so inter-vortex interactions have decayed to zero. Then the equation of motion becomes a force balance between the viscous drag term and the sum over repulsive vortex-vortex interactions, roughly over a penetration depth area. Replacing the discrete sum over vortices with an integral over density gives an equation of the form

\begin{equation}
\eta {\bf v}({\bf r}) = \int {\rm d} {\bf r}' f^{vv}({\bf r} - {\bf r}') \rho({\bf r}') \widehat{({\bf r}-{\bf r}')}
\label{eq:forcebalance}
\end{equation}

The viscous term on the left hand side is due to the ``normal fluid" of excited quasiparticles scattering from the vortex, or trapped in its core.  We now use the small value for the change in vortex density over a distance of the penetration depth to approximate \Eqref{eq:forcebalance}. Performing a change of basis ${\bf r}' \rightarrow {\bf r} +{\boldsymbol \zeta}$ and Taylor expanding $\rho({\bf r}+\boldsymbol\zeta)$ to first order in $\boldsymbol\zeta$ gives the transformed equation

\begin{equation}
\eta {\bf v}({\bf r}) \simeq \int {\rm d} {\boldsymbol \zeta} f^{vv}({\boldsymbol \zeta}) \left[\rho({\bf r}) + {\boldsymbol \zeta}\cdot \nabla \rho({\bf r})\right] \widehat{{\boldsymbol\zeta}}
\end{equation}

Since ${\boldsymbol \zeta} = (\zeta \cos \phi, \zeta \sin \phi )$, $\widehat{{\boldsymbol \zeta}} = ( \cos \phi, \sin \phi) $  and $\nabla\rho= (\partial \rho / \partial x, \partial \rho / \partial y )$, only the term in $\partial \rho / \partial x$ survives, which results in

\begin{equation}
v(x) = - \frac{\Phi_0^2}{\eta \mu_0} \frac{ {\rm d} \rho}{ {\rm d} x}
\label{eq:velx}
\end{equation}

To determine the density profile the steady state continuity equation is introduced, in 1D this is $Q = \rho(x) v(x)$. Substituting \Eqref{eq:velx} and performing the simple integration gives the density profile

\begin{equation}
\rho(x) = \sqrt{\rho(0)^2 - \frac{\eta Q  \mu_0}{\Phi_0^2}\thinspace x}
\label{eq:rhox}
\end{equation}

where the boundary condition on the entrance to the channel $\rho=\rho(0)$ was used.

\end{document}